\documentclass[letterpaper]{article}
\usepackage{times}  
\usepackage{helvet}  
\usepackage{courier}  
\usepackage[hyphens]{url}  
\usepackage{graphicx} 
\usepackage{caption}

\usepackage{mathtools}
\usepackage{bm}

\usepackage[style=numeric,citestyle=numeric-comp , giveninits=true, backend=biber]{biblatex}

\usepackage{xcolor}

\usepackage{amsfonts}       
\usepackage{amsmath}
\usepackage{dsfont}

\newcommand\Ea{E_{\text{QE}}}
\newcommand\J{{\text{J}}}
\newcommand{\Id}{\mathds{1}}
\newcommand{\bb}{\bm}
\renewcommand{\L}{\mathcal L}
\renewcommand{\d}{\text{d}}
\newcommand{\Jh}{\text{J}_h} 

\title{Neural Learning Rules from\\ Associative Networks Theory}
\date{}

\author{%
  Daniele Lotito \\
IBM Research -- Europe \footnote{Work done during an Internship at IBM Research -- Europe while the author was affiliated with University of Pisa}\\
  \texttt{daniele.lotito@ibm.com}\\
  \hspace{100mm}\\
  }

\begin{document}

\maketitle

\begin{abstract}
Associative networks theory is increasingly providing tools to interpret update rules of artificial neural networks. At the same time, deriving neural learning rules from a solid theory remains a fundamental challenge. \\
We make some steps in this direction by considering general energy-based associative networks of continuous neurons and synapses that evolve in multiple time scales. We use the separation of these timescales to recover a limit in which the activation of the neurons, the energy of the system and the neural dynamics can all be recovered from a generating function.  
By allowing the generating function to depend on memories, we recover the conventional Hebbian modeling choice for the interaction strength between neurons.
Finally, we propose and discuss a dynamics of memories that enables us to include learning in this framework. 
\end{abstract}

\section{Introduction}
Associative learning and energy-based modeling are central paradigms in machine learning. At their intersection lies the central problem of designing an energy functional that leads to a well behaved and efficient network dynamics, and associative networks theory provides us with valuable principles to set up an attractor dynamics 
 with the desirable properties.
A significant contribution that comes from the study of associative memories is the equivalence of the attention mechanism in transformers~\cite{Vaswani2017} with the updating rule of an associative memory network~\cite{ramsauer2021hopfield} strictly related to the Hopfield model~\cite{  hopfield1982, Hopfield1984}. \\
This encouraging result led to a broad interest in the application of spin glass models~\cite{MPV} to the Artificial Intelligence domain, which resulted in the investigation of a new class of models that display promising computational capabilities, such as the possibility to store and retrieve a number of memories that is proportional to the number of neurons raised to the degree of interaction; this could constitute a new frontier for Hopfield networks~\cite{krotov2023new}.
Although this investigation was at the beginning primarily analytical, with first steps being analyzing the computational properties of dense associative networks~\cite{ dense_sup,dense_unsup} to exploit them in machine learning settings~\cite{barra2012equivalence, Barra2018, Agliari2022} and build theoretical groundings for AI models~\cite{Krotov2016,Krotov2018, hoover2023memory, ambrogioni2023search, schaeffer2024bridging, krotov2021hierarchical}, more recently it has been understood that large scale models based on associative networks achieve satisfactory performances that make them suitable for practical use cases~\cite{liang2022modern,  kozachkov2023neuronastrocyte,hoover2024energy, widrich2020}.

An important common ground of associative networks and machine learning is the description of the dynamics of their computing units, that are represented by neurons and synapses. These learning rules should be computationally viable, for instance, the attractor-based dynamics in associative networks poses serious computational challenges~\cite{dense_unsup}. 
In biological networks modeling these learning rules should also be biologically plausible, for instance, in synaptic modeling, we cannot directly model synapses as tensors of order greater than three, thus the many-body interaction terms of dense associative memories do not fit this purpose apparently.\\

A significant step forward has been recently made by Dmitry Krotov, John Hopfield, and collaborators. It consists in the development of a general framework for artificial neural networks that directly descends from associative learning and energy-based model theory~\cite{krotov2021large, hoover2022a}.

In a nutshell, Krotov and collaborators consider a network of fast (visible) neurons connected with (slow) hidden neurons. Under the assumption of well-separated time scales, the network dynamics can be derived from a convex function. The Legendre transformation of this generating function can be interpreted as the network energy as it is minimized by its dynamics. A more detailed discussion of Krotov's work can be found in Sec.~\ref{sec:discussion}.

Krotov was able to find generating functions that allow the recovery of neural update rules of famous models such as modern Hopfield networks~\cite{krotov2023new} and attention mechanism~\cite{Vaswani2017, Ramsauer2021}. \\
However, his work lacks some important features, which we are about to highlight. We address these issues in this paper.

Krotov introduced a memory matrix whose elements correspond to the interaction strengths between neurons that evolve to minimize the network's energy. In traditional associative network literature, however, the interaction strength between pairs of neurons is determined by the matrix product of the memory matrix and its transpose; this product is known as the Hebbian matrix. We show that Hebbian theory is recovered using a generating function dependent on the memory matrix.

Moreover, memories do not evolve in Krotov's framework, which means we cannot learn new memories or update stored ones. To use this scheme to model learning, we need to introduce a dynamic for memories. \\
We propose and theoretically study the dynamics of these parameters and show that is compatible to Hebbian learning~\cite{Agliari2023, lotito2024learningassociativenetworkspavlovian}.

\subsection{Main Results}\label{sec:results}
Our ultimate aim is to advance toward a unified perspective on learning in associative neural networks.
We take a step in this direction by presenting the following advancements:
\begin{itemize}
    \item \textbf{Extending the framework by Krotov et al.}~\cite{krotov2021large, hoover2022a}: We introduce two key features:
    \begin{enumerate}
        \item Generating functions that depend on the synaptic matrix, discussed in Sections~\ref{sec:memory_generating} and~\ref{sec:memory_generating_example}.
        \item Synaptic evolution mechanisms, elaborated in Sec.~\ref{sec:hierarchical}.
    \end{enumerate}
    \item \textbf{Connecting to other research fields:} Inspired by recent efforts in linking associative networks with probabilistic modeling~\cite{schaeffer2024bridging, NEURIPS2023_4e8a7498}, we establish connections with:
    \begin{enumerate}
        \item Hebbian learning theory, which forms the foundation of the Hopfield model and its extensions (Sec.~\ref{sec:Synaptic_discussion}). This important connection provides the theoretical justification for our proposed extensions.
        \item Results random matrix theory and dynamical systems theory (see Sec.~\ref{sec:connection_dyn} and  Appendix~\ref{app:HG}).
    \end{enumerate}
\end{itemize}
We believe that associative network theory holds significant potential for modeling learning mechanisms, at the same time, various branches of mathematics are linked to it. We hope to foster the application of mathematics to associative network theory. For this reason, we provide the mathematical foundation of Krotov's take on associative networks, and we present our results on this theoretical ground, see section ~\ref{sec:theoretical} for further discussion on this point.

\subsection{Structure of the paper}\label{sec:structure}
In order to reach our goals, see Sec.~\ref{sec:results}, we introduce the modeling tools, show that they are sufficient to encompass Krotov's work, and then discuss its generalization and its connection to various mathematical disciplines.
We begin by presenting the intuition behind the theory in Sec.~\ref{sec:preamble}, where we also explain why this framework is well-suited for describing machine learning architectures. 
Next, in Sec.~\ref{sec:energy}, we shift to a more formal mathematical formulation of Krotov's work. In this section, we provide a general description of the Lagrange transformation mechanism, which lays the foundation for the new results introduced from Sec.~\ref{sec:memory_generating}.\\
In section~\ref{sec:hierarchical} we highlight that the model results from the quasi-equilibrium approximation of a multiple timescale system of non-linear partial differential equations. In this section, we discuss the synaptic dynamics, a fundamental component to be able to describe learning within this framework. \\
In the discussion section, Sec.~\ref{sec:discussion} we highlight the connections with various disciplines, such as random matrix theory, dynamical systems theory, statistical physics, and associative network theory.
Additionally, we provide examples that demonstrate the necessity of the generalizations introduced to describe other associative network models within a unified theoretical framework.\\
Finally, we conclude our work in section~\ref{sec:conclusions}.

\section{Modeling assumptions}\label{sec:preamble}
The system under our consideration is a network of feature neurons, hidden neurons and synapses,  both neurons and synapses can evolve, and their dynamics happens on different timescales. \\
These components can be arranged in multiple layer following the theory developed in~\cite{krotov2021hierarchical}; we will use a similar notation of the closely related paper~\cite{krotov2021large}.\\
First of all, let us consider a network of $N_v$ visible neurons $\bb v \equiv \{v_i\}_{i = 1,\dots, N_v}$, and $N_h$ memory neurons $\bb h \equiv \{h_\mu\}_{\mu = 1,\dots, N_h}$, we denote $N=N_v +N_h$ the total number of neurons and use $\{\bb x\}_{i = 1,\dots, N}$ when we refer to both kinds of neurons; we defer the description of synapses to section~\ref{sec:hierarchical}.
As associative networks follow an attractor dynamics, each of our neurons is attracted towards an attractor that depends upon the other components of the network. We assume that this dynamics follows an exponential decay,\footnote{Investigating the effects of other kinds of decay is left for future work. Here we stick to the widest spread practice in associative networks theory, see e.g. \cite{Hopfield1984, krotov2021large, huang2022extreme}.}
\begin{equation}\label{eq:exp_decay_general}
  \tau_x \dot x_i = B_i - x_i 
\end{equation}
with $B_i$ a fixed scalar, and $\tau_x$ being the typical timescale of the variable under consideration. We set $\tau_x = \tau_v$ for all the neurons in the visible layers, while $\tau_x = \tau_h$ for those in the hidden layer.\\
The visible neurons layer contains variables that are easier to observe, since they vary slower; in other words, we assume $\tau_v > \tau_h$. \\
We will introduce non trivial dynamics by allowing $B_i$ to be a function of the network parameters. \\
We immediately notice that~\eqref{eq:exp_decay_general} is the continuous counterpart of a neural updating rule. Writing equation~\eqref{eq:exp_decay_general} in finite differences gives
\begin{equation}\label{eq:finite_diff}
  x_i ^{(t+1)} = x_i^{(t)} + \frac{\d t}{\tau_x}\left( B_i - x_i^{(t)} \right) , 
\end{equation}
which for $dt  = \tau_x$ becomes $x_i ^{(t+1)} = B_i$. \\
Thus, the network reaches the state $\bb B$ in a single update. Single step convergence has been found elsewhere when modeling deep learning architectures using associative network theory~\cite{ramsauer2021hopfield}.
The choice $dt  = \tau_x$ is completely arbitrary, as $dt$ corresponds to the discretionary characteristic time of the updating step and $\tau_x$ is determined within the model.
With similar considerations, in~\cite{krotov2021large} is illustrated how a simple dynamics such as~\eqref{eq:exp_decay_general} can recover the update rule of~\cite{ramsauer2021hopfield}, while applying it once corresponds to perform the dot-product attention used in transformers networks~\cite{Vaswani2017}. 
Of course, in their analysis $\bb B$ is not a vector of fixed real numbers. \\
Now, we relax this assumption on $\bb B$.
Generally,  for the update step $(t+1)$, $B_i =B_i^{(t)}$ and $B_i ^{(t)} =B_i ^{(t)} ( \{\bb x ^{(\tau)}\}_{\tau=1,\dots, t} ; \{\bb \xi\} )$, where $\{\bb \xi\}$ is the set of $N_h$ memories of the network, that are vectors of $N_v$ elements; these memories determine how pairs of neurons interact.  \\
We assume no infra-layer connections, as usual in machine learning. Thus, we consider a $[N_h \times N_v]$ matrix $\Xi$, that we call \textit{memory matrix}, the row of this matrix being individual memories $\xi ^ \mu$, $\mu = 1, \dots , N_h$. 
In principle, this matrix can change too, we denote its typical time scale $\tau_\Xi$, and assume straight away that $\tau_\Xi \gg \tau_v$. This assumption implies that $\Xi$ can be seen as a matrix of parameters through the whole dynamics. \\
Learning coupling matrices is a central problem in both associative network theory and energy based modeling, and it is usually accomplished through the use of gradient descent schemas~\cite{schaeffer2024bridging}. Treating synaptic variables in the same fashion as neural variables, with the difference that they evolve in a wider time scale allows for a unified description of the whole network, and this idea has been poorly explored, see e.g.~\cite{lotito2024learningassociativenetworkspavlovian, Agliari2023}. Including a description of synaptic dynamics in Krotov's picture is one of the contributions of this work. \\
In section~\ref{sec:hierarchical} we will discuss the evolution of memories within our framework.\\  
We further assume that $\bb B ^{(t)}$ does not depend on previous updates, $B_i ^{(t)} =B_i ^{(t)} (\{\bb x ^{(t)}\} ; \Xi )$, this makes the whole dynamics Markovian. This is not an essential assumption of our framework, but we leave the exploration of the non-Markovian setting for a future work. \\ 
To chose the specific form of $\bb B^{(t)}$ we need to further develop our theory.

\section{Energy-based modeling}\label{sec:energy}
We focus now on the case in which the three timescales are well separated, $\tau_\Xi \gg \tau _v \gg \tau _h$, we call this scenario \textit{quasi-equilibrium}, as customary in multiple time scale modeling~\cite{verhulst2005methods, GRASMAN197893}. \\
In this case, we are able to write the energy of our model through the Legendre transformation of a convex generating function $\L \in \mathcal C ^2$.
\footnote{Here, we only point out that $\L$ is called Lagrangian in~\cite{krotov2021large} and related literature by virtue of the similitude with classical mechanics, we find this nomenclature potentially misleading. For instance, only in the quasi-equilibrium scenario, it is possible to straightforwardly obtain the energy by Legendre-transforming $\L$} \\ 
We consider the visible neurons timescale. In this case, the hidden neurons reach equilibrium instantaneously, since $\tau _ h \rightarrow 0$. This is the quasi-equilibrium approximation and sets the value of hidden neurons equal to $\bb h = \bb h (\bb v)$.

The generating function can be written as
\[ \L = \L_ v  + \L _ h     ,       \]
where $\nabla _{\bb v} \L_h = \bb 0 $, $\nabla _{\bb h} \L_v = \bb 0 $ and $\L = \L(\bb v , \bb  h)$.\\ We can now Legendre transform on the $\bb v$ coordinates to obtain
\begin{equation}\label{eq:EQE}
  \Ea = \sum _i ^{N_v} (v_i - I_i ) \frac{\partial \L }{\partial (v_i - I_i)} - \L ,
\end{equation}
where we added a constant field contribution to each neuron, in many cases $I_i = 0$ for every neuron.
The Legendre transformation changes the set of independent variables, from $\{\bb v, \bb h \}$ to $\{ \bb g , \bb h \}$, where $\bb g \equiv \nabla _{\bb v} \L = \nabla _{\bb v} \L _v = \nabla _{(\bb v - \bb I)} \L _v$. Since $\bb h = \bb h (\bb v)$ in the quasi-equilibrium approximation, we also have $\Ea = \Ea ( \bb g  )$.\\

The differential of the quasi-equilibrium energy reads as
\begin{equation}\label{eq:eadi_d}
  \d \Ea = \d \bb g \cdot (\bb v - \bb I)  - \d \bb g \cdot \Jh \, \nabla^{\top} _{\bb h} \L _h \, .
\end{equation} 
Where $\Jh = \nabla _{\bb g} \bb h$ is the Jacobian matrix, whose dimensions are $[N_v \times N_h]$, and $\nabla^{\top} (\, \cdot \, ) = [ \nabla (\, \cdot \, )] ^{\top}$.\\
We observe that $\bb g$ is a vector of functions that assigns to a given vector, the gradient $\nabla _{\bb v} \L _v$ evaluated at that vector,  when this vector is not specified it is intended to be $\bb v$. Analogously,  to shorten the notation, we define $\bb f \equiv \nabla _{\bb h} \L _h$; we now compute 
\begin{equation}\label{eq:eadi_dt}
  \frac{ \text{d}}{  \text{d} t} \Ea(t) = \dot {\bb g}  \cdot  \nabla _{\bb g} \Ea = \dot {\bb g}  \cdot \left( (\bb v - \bb I) -  \Jh \, \bb f ^{\top}\right) .
\end{equation}
Re-establishing the dependence of $\bb g$ on $\bb v$ we arrive at 
\begin{equation}\label{eq:e_decreasing}
\frac{ \text{d}}{  \text{d} t} \Ea(t) = -\tau _v \dot {\bb v}^\top \mathbf {H}_v \dot {\bb v} ,
\end{equation}
where we have noticed that constraining the dynamics of the visible neurons to 
\begin{equation}\label{eq:vdyn}
  \tau _v \dot {\bb v} = - \bb v + \bb I + \Jh \, \bb f ^{\top} ,
\end{equation}
is equivalent to assuming~\eqref{eq:exp_decay_general}. Moreover, $\mathbf {H}_v $ corresponds to the Hessian of $\L _v$. 
The convexity of the generating function leads to a positive definite Hessian, then
\begin{equation}
 \frac{ \text{d}}{  \text{d} t} \Ea(t) < 0 .
\end{equation}
This proves that under our assumptions the energy-based model is well behaved, since the energy decreases along the dynamical trajectory~\eqref{eq:vdyn}. Furthermore,  $\partial _ t \Ea(t) = 0$ if and only if we reach a fixed point of the dynamics, $\dot {\bb v} = 0$.\\
\subsection{The hidden neurons algebraic equation}
After having illustrated the general approach, we now discuss the main choices for the hidden neurons algebraic equation $\bb h - \bb h (\bb v) = 0$.
We remark that, assuming the exponential decay~\eqref{eq:exp_decay_general}, this is the equilibrium solution of the evolution $\tau_h \dot{\bb h} = - \bb h + \bb h (\bb v)$.\\ 
An interesting case arises when $\Jh$ is a constant matrix $\text{J}$, 
the hidden neurons algebraic equation is of the form $\bb h = \text{J} \nabla_{\bb v}^{\top} \L  + \bb b$, where $\bb b$ is a fixed vector of biases that corresponds to the integration constant obtained by solving $\text{J} = \nabla _{\bb g} \bb h$ with respect to $\bb h$. \\
Thus, $\bb h$ stores the dot products between the intra-layer couplings and the gradient $\nabla_{\bb v} \L  $, plus a bias vector. 
For this reason, we interpret $\partial _{v_i} \L$ as the activation function of $v_i$. \\
Setting $\J = \Id$ accounts for reducing the neural space to $\mathbb{R} ^m \times \mathbb{R} ^m$, where  $m = {\min \{N_h, N_v \}}$, and completely relying on the generating function to produce non trivial dynamics.\\
While, in the case $\J = \Xi$, the visible neuron dynamics is
\begin{equation}\label{eq:vxidyn}
  \tau _v \dot{\bb v} = - \bb v +\bb I + \Xi^\top \bb f( \Xi\bb g ^{\top}  + \bb b ) ^ \top  \, .
\end{equation}
This form of the visible neurons dynamics recovers the dynamics of~\cite{krotov2021large}.
In that paper it has been observed that the update rule of transformer networks~\cite{Vaswani2017} can be obtained from a dynamics such as~\eqref{eq:vxidyn} discretized as~\eqref{eq:finite_diff}.
To sum up, under the quasi-equilibrium approximation for slow–fast dynamical systems~\cite{verhulst2005methods, GRASMAN197893}, the system can be derived from a single convex generating function. \\
We have thus far presented Krotov's framework in a mathematical context; the remainder of this paper focuses on introducing new results.
\subsection{The coupling matrix and the memory matrix}\label{sec:memory_generating}
The role of the coupling matrix $\J$ is to linearly transform objects in the hidden neurons space to the visible neurons one, $\J \in \mathbb{R} ^{N_h} \times \mathbb{R} ^{N_v}$, note that it has the same dimensions as the memory matrix.\\
In this work, we consider either $\J = \Xi$ or $\J = \Id$, where $\Id$ is the extended identity matrix; this is motivated by the fact that we want all network parameters to be stored in the memory matrix, and, for the same reason, we allow the generating function $\L$ to depend on the memory matrix as well. Krotov and coauthors do not allow the generating function to be dependent on the memory matrix ~\cite{krotov2021large, hoover2022a}. We explain in the discussion section, namely Sec.~\ref{sec:memory_generating_example},  that this generalization can be a useful modeling tool that can also be biologically plausible. We also provide a use case for the case $\J = \Id$  and $\L= \L(\Xi)$.\\

\section{Multiple time scale modeling}\label{sec:hierarchical}
Under the assumption of well-separated timescales both the network energy and the neurons activations derive from the same generating function.
Separating the neural timescales is equivalent to \textit{integrating out} the contribution of hidden neurons, that no longer interact meaningfully with visible neurons, but rather are determined by their state.\\
The idea of integrating out hidden neurons to recover an effective dynamics for feature neurons is not new, it is e.g. at the core of Restricted Boltzmann Machines \cite{smolensky1986information, Hinton2012a}, whose dynamics can be recovered with Hopfield networks~\cite{barra2012equivalence}, and has been discussed also in the related work~\cite{krotov2021large}.
It results in considering separate subsystems, the slow system and the fast system. The full system evolves in a subspace where the dynamical variables of the slow system are determined by an algebraic equation such as $\bb h = \J\nabla_{\bb v}^{\top} \L  + \bb b$. In our case, the state of the network is determined by solving a differential algebraic equation system, the case in which the separation of the timescales is not complete will be a topic for a future work; here we mention that these fast--slow dynamical systems are ubiquitous in science~\cite{rinzel1985, izhikevich2007dynamical, hek2010geometric, weinan2011principles, bertram2017multi}. The study of their multiscale structure~\cite{weinan2011principles, pavliotis2008multiscale, witelski2015methods, verhulst2005methods, GRASMAN197893} can benefit from geometric singular perturbation theory \cite{fenichel1979geometric, Jones1995, wechselberger2020geometric}.

We zoom out to grasp the bigger picture, the neural network we are considering is a multiscale system of partial differential equations.
The equation for the neural dynamics are
\begin{equation}\label{eq:neuralsys}
  \begin{cases}
    \tau _v \dot {\bb v} = - \bb v + \bb I + \Jh \, \bb f ^{\top} \\
      \tau _h \dot {\bb h} = - \bb h + \bb h(\bb v) \, . 
  \end{cases}
\end{equation}
The quasi-equilibrium approximation leads to the algebraic differential equation system
\begin{equation}\label{eq:DAE}
  \begin{cases}
    \tau _v \dot {\bb v} = - \bb v + \bb I + \Jh \, \bb f ^{\top}\\
      {\bb h} = \bb h (\bb v)
  \end{cases}
\end{equation}
that we have discussed in the previous section.\\

We now move to the update rule for the memory matrix, in the subsection~\ref{sec:Hop} we discuss the connection with the Hebbian prescription for Hopfield-like models.\\
The memory dynamics is governed by the following differential equation:
\begin{equation}\label{eq:memdyn}
  \tau_\Xi \dot{\Xi} = \bb w \odot \left[ (-\Xi + \Lambda ) \bb{+} \bb v^\top \right]\, .
\end{equation}
In this expression, $\odot$ represents the Hadamard (element-wise) product between the weight vector \(\bb w\) and the matrix. The row vector $\bb v ^\top$ is broadcasted across rows indexed by $\mu = 1, \dots, N_h$.\\
The weight vector \(\bb w\) is obtained through the softmax function:
\begin{equation}
  w_\mu = \frac{e^{\beta (\Xi \bb v)_\mu}}{\sum_{\nu=1}^{N_h} e^{\beta (\Xi \bb v)_\nu}}.
  \label{eq:softmax}
\end{equation}
The parameter \(\beta\) determines the selectivity of the weighting: as \(\beta \to \infty\), \(\bb w\) converges to a one-hot vector that selects the row of \(\Xi\) exhibiting maximal similarity with the visible state \(\bb v\). The Hadamard product with \(\bb w\) (particularly for large \(\beta\), which we hereafter assume throughout this work) ensures that updates primarily affect the selected row while minimizing changes to other rows. The synaptic bias matrix \(\Lambda\), which shares dimensions with \(\Xi\), serves a role analogous to the bias vectors \(\bb b\) and \(\bb I\).\\
The evolution equation \eqref{eq:memdyn} can be expressed element-wise as:
\begin{align}
  \tau_\Xi \dot{\Xi}_{\mu i} &= w_\mu \left[ (-\Xi_{\mu i} + \Lambda_{\mu i}) + v_i \right] \notag \\
  &= \frac{e^{\beta (\Xi_\mu  \cdot \bb v)}}{\sum_{\nu=1}^{N_h} e^{\beta (\Xi_\nu \cdot \bb v)}} \left[ (-\Xi_{\mu i} + \Lambda_{\mu i}) + v_i \right]
  \label{eq:memdyn_el}
\end{align}
where $\Xi_\mu$ represents row $\mu$ of $\Xi$.
Assuming zero bias \(\Lambda = 0\), we can easily imagine to start from a set of old memories \(\{ \bb \xi^{(0)} \}\) that form the memory matrix \(\Xi^{(0)}\). We update these memories based on observed patterns \(\{ \bb \xi^\prime \}\), which correspond to the values assumed by visible neurons and can be grouped into a new memory matrix \(\Xi^\prime\).\\
Typically, to mimic a learning process, the network ``sees'' through visible neurons $\bb v$ that are ``clamped'' to patterns in \(\{ \bb \xi^\prime \}\). For example, for $0<t<t_1$ we have  $\bb v (t)= \bb \xi^\prime _{\mu_1}$, for $t_1<t<t_2$,  $\bb v (t)= \bb \xi^\prime _{\mu_2}$, and so on. The sequence of indexes $\mu_1, \, \mu_2, \dots$ is obtained through a stochastic process so that
\begin{enumerate}
\item  At each time $t$, \(\bb v(t) = \bb \xi^\prime_\mu\) with probability \(p_\mu\), where \(\{p_\mu\}_{\mu = 1, \dots, N_h}\) satisfies \(p_\mu \geq 0\) and \(\sum_\mu^{N_h} p_\mu = 1\).
\item  The clamping events are independent.
\end{enumerate}
During each time interval, rows of the coupling matrix that are more similar to the visible neuron configuration decay more rapidly toward it. The probabilities $\{ p_\mu\}_{\mu = 1, \dots, N_h}$ further tune this decay, making the products $\{ \frac{p_\mu w_\mu}{\tau _\Xi} \doteq \tau_\mu \}_{\mu = 1, \dots, N_h}$ behave as the relevant time scales for each pattern, this is an ansatz that we corroborate in Appendix~\ref{app:poc}.\\
To ensure proper initialization of the learning process, reasonable assumptions are that each pattern \( \bb \xi^\prime \) is a perturbed version of its corresponding memory \( \bb \xi^{(0)} \) and that memories are well distinguishable from one another (we remand again to Appendix~\ref{app:poc} for details). This condition can be expressed as
\begin{equation}\label{eq:condition}
  \frac{\bb \xi^{(0)}_\mu \cdot \bb \xi^{(0)}_\nu}{\| \bb \xi^{(0)}_\mu \| \| \bb \xi^{(0)}_\nu \|} \approx \delta_{\mu \nu} 
\end{equation}
for each \(\mu, \nu \in \{1, \dots, N_h\}\). \\
The goal of synaptic dynamics is for each memory row to converge to its corresponding updated pattern:  
\begin{equation}\label{eq:conv_mem}  
  \Xi^{(0)} \longrightarrow \Xi^{(\infty)} = \Xi^\prime.  
\end{equation}  
In the limit $\beta \to \infty$ this convergence is trivially obtained, provided that $p_\mu \neq 0$ for each $\mu$, since only one row is updated at the time.\\
A detailed analysis of synaptic dynamics across different settings is beyond the scope of this work. However, in Appendix~\ref{app:poc}, as a proof of concept, we numerically verify the convergence stated in Eq.~\ref{eq:conv_mem} within a typical statistical mechanics framework.
\section{Discussion}\label{sec:discussion}
We have rephrased Krotov's picture of associative networks on a solid theoretical ground and then extended it in two directions: we have allowed the generating function to depend on the memory matrix, and we have proposed a dynamics for the synaptic matrix.\\
Now, we elaborate on these three points.
\subsection{Theoretical Takeaways}\label{sec:theoretical}
We examine here the advantages of our theoretical formulation of Krotov's model. We consider the case of modern Hopfield networks as an illustrative example.\\
Krotov assumes that the dynamics of each neural family ($ \bb v$, $\bb h$) is linear in the outputs of the other family ($\bb f $, $\bb g)$.
\begin{equation}\label{eq:krotovsys}
    \begin{cases}
        \tau _v \dot v _i  = \sum_{\mu = 1}^{N_h} \xi  _{i \mu } f_\mu - v_i + I_i \, ,\\
        \tau _h \dot  h _ \mu = \sum_{i = 1}^{N_v} \xi _{\mu i } g _ i - h_\mu \, .
    \end{cases}
\end{equation}
This dynamics can be obtained by taking the gradients 
\begin{equation}\label{eq:kdyn}
  \nabla _{\bb g} E = - \tau _v \dot {\bb v}   \, \ \ \ \text{and} \ \ \  \nabla _{\bb f} E= -  \tau _h \dot {\bb h }
\end{equation}
of the energy 
\begin{equation}\label{eq:Ken} 
    E= \left[ \sum_{i = 1}^{N_v}(v_i - I_i ) g_i - \mathcal L _v \right]  + \left[ \sum_{\mu = 1}^{N_h} h_\mu f_\mu - \mathcal L _h \right] - \sum _{i,\mu} f_\mu \xi _ {\mu i } g_i .
\end{equation}
Here, $\mathcal{L}_v$ and $\mathcal{L}_h$ are chosen such that $g_i = \frac{\partial \mathcal{L}_v}{\partial v_i}$ and $f_\mu = \frac{\partial \mathcal{L}_h}{\partial h_\mu}$.
Krotov's energy function \eqref{eq:Ken} is thus composed of the Legendre transformations of the neural generating function $\L = \L_v + \L_h$ and of the last term that has been added by hand. This is to ensure that in the quasi-equilibrium limit  \(\tau_h \to 0\), it holds that $h_\mu = \sum_{i = 1}^{N_v} \xi_{\mu i} g_i $, so that the first term in the second bracket of  \eqref{eq:Ken} vanishes. Krotov's then restricts the discussion to the slow variables only.

We focus on the quasi-equilibrium regime highlighting that it is characterized by a far richer structure than previously noticed. The core feature of this treatment is that the fast variables are constrained to the slow one through the arbitrary function $\mathbf{h} = \mathbf{h}(\mathbf{v})$. \\
This brings two advantages:
\begin{itemize}
    \item First, by considering the quasi-equilibrium limit from the beginning we avoid adding by hand the term $\sum _{i,\mu} f_\mu \xi _ {\mu i } g_i$. We focus on slow variables $\bb v$ and derive the quasi-equilibrium energy function directly through the Legendre transformation of the generating function with respect to these variables only, see Eq.~\ref{eq:EQE}, reported here
    \begin{equation*}\label{eq:Ken_2} 
      E_{QE}=  \sum_{i = 1}^{N_v}(v_i - I_i ) g_i - (\mathcal L _v  + \mathcal L _h ) \ .
    \end{equation*}
    \item Second, we allow for more general learning rules than those in system~\eqref{eq:krotovsys}. Taking as a reference Eq.~\ref{eq:vdyn}, the gradient $\mathbf{J}_h = \nabla{_\mathbf{g}} \mathbf{h}$ can still depend on $\mathbf{h} = \mathbf{h}(\mathbf{v})$. While this means we lose the interpretation that each neuron's input is a linear combination of outputs from neurons in the other family, the ${\mathbf{h}}$ variables can now be interpreted as latent fast variables -- not necessarily neurons.
These variables can model biological components (e.g., glial cells \cite{LINNE2024102838, kozachkov2023neuronastrocyte}) or serve as latent behavioral variables that, while not directly observable, can be inferred through mathematical models of cognitive processes~\cite{musall2019harnessing}. A successful example of the latter is the inferred accumulation of sensory evidence in decision-making studies~\cite{shadlen2013decision}. The exploration of these connections will be pursued in future work.

\end{itemize}

\subsection{Generating functions that depend on the memory matrix}\label{sec:memory_generating_example}
Here, we discuss the implications of having a generating function that depends on the memory matrix.
In this theory, as we anticipated in Sec.~\ref{sec:memory_generating}, we want all parameters to be stored in the matrix $\Xi$. After the discussion of Sec.~\ref{sec:hierarchical} we can appreciate the meaning of this choice: these parameters can now be updated through a learning mechanism, ensuring that our modeling framework retains this flexibility. \\

First, this extension does not necessarily contradict the definition of ``biological plausibility'' adopted in~\cite{krotov2021large} where it is assumed to correspond to the absence of many-body synapses. 
It is worth noting that an alternative approach to reconciling many-body synapses with biological interpretation lies in the use of $q$-state Potts networks, in which each neuron can assume $q$ different discrete states~\cite{kanter1988potts}. Notably, similar to how many-body interactions between slow neurons emerge from an effective description of a well-separated slow-fast system~\cite{krotov2021large},  many-body interactions between binary neurons in Ising networks can, under certain conditions, be effectively represented using Potts neurons interacting in pairs~\cite{Kanter_1987}. \\
Another important point is that in~\cite{krotov2021large}, the memory patterns $\bb \xi_\mu$ are interpreted as the strengths of the synapses connecting slow and fast neurons. This interpretation is different from the conventional interpretation, in which the strengths of the synapses are determined by matrices which are either outer products of the memory vectors or higher-order generalizations of outer products~\cite{krotov2021large}.

The conventional interpretation is the one that gave associative network theory its modeling power in the biological domain. 
Extending Krotov's framework to be able to recover this interpretation is thus fundamental. \\
We see that by letting the generating function depend on the memory matrix, we can recover the conventional interpretation through the following example. 
\subsubsection*{Example: memory-dependent generating function}
We adopt the choices $\J = \Id$, and $\L = \L(\bb v, \bb h; \, \Xi)$. We consider a system in which $N_v = N_h$ and $\bb I = \bb 0$, in this case~\eqref{eq:vdyn} becomes

\begin{equation}\label{eq:vdyn_id}
  \tau _v \dot {\bb v} = - \bb v + \bb f(  \bb g ^{\top}  + \bb b ) ^ \top    ,
\end{equation}

Let us consider $\L _h = \sum_ \mu \ln (\cosh h _ \mu )$, $\L _v = \frac{\tilde \beta }{2} \, \| \Xi \bb v\| ^2$, with the norm $\| \Xi \bb v\| ^2 = \bb v ^ \top \Xi ^ \top  \Xi \bb v$, and $\tilde{\beta}$ being a constant to be set afterward. \\
The generating functions are well behaved;
 since $\partial _x \ln (\cosh x) = \tanh (x)$ and $\nabla _{\bb v} \L_v = \tilde \beta \, \Xi ^\top \Xi \bb \, \bb v$, we can write the dynamics~\eqref{eq:vdyn} as 
\begin{equation}\label{eq:almost_Hebb}
  \tau _ v \dot {\bb v} = - \bb v + \tanh ( \tilde \beta \, \Xi ^\top \Xi \bb \, \bb v  ^\top + \bb b) , 
\end{equation}
where the hyperbolic tangent acts element-wise.\\
This neural learning rule corresponds to the dynamics 
  \(  \tau _v \dot {\bb  v} = - \bb v + \tanh(\beta \bb J \bb  v^ \top + \beta \bb u  ),  \)
that has been proposed to model classical conditioning with associative networks~\cite{lotito2024learningassociativenetworkspavlovian, Agliari2023}. 
To unveil the equivalence it is sufficient to identify $\bb b= \beta \bb u $, $\tilde{\beta} = \frac{\beta}{N_h}$, and $\bb J = \frac{1}{N_h} \, \Xi ^\top \Xi$. \\
We recall that memories correspond to the (transposed) rows of the memory matrix $\bb \xi ^\mu = (\Xi _{\mu \bb\cdot})^\top$.
Thus, the matrix elements are given by $J_{ij} = \frac{1}{N_h}\sum _\mu \xi ^\mu _i \xi ^\mu _j $.\\
In this way we have recovered the so-called Hebbian prescription~\cite{AGS_prl}.\footnote{
We mention that in~\cite{lotito2024learningassociativenetworkspavlovian, Agliari2023} self-interaction terms are not considered ab initio, thus their update rule results in \(
  \tau _v \dot v_i = - v + \tanh(\beta \sum_{j\neq i} J_{ij} v_j + b_i)  \). We discuss connections with Hebbian theory also in Sec.~\ref{sec:Synaptic_discussion}.}
We thus recover the conventional interpretation for neural couplings, since the matrix multiplying the neural vector $\bb v$ is the outer product of memory vector, $\Xi ^\top \Xi$.\\
We also observe that the evolution~\eqref{eq:almost_Hebb} differs from the graded response Hopfield model~\cite{Hopfield1984}, 
 \(\tau _v \dot {\bb  v} = - \bb v + \bb J \tanh(\beta  \bb  v^ \top + \beta \bb u  ),  \) only by the fact that the multiplication by the coupling matrix is carried inside the non-linearity (here, the hyperbolic tangent acts element-wise). \\
Applications of the quasi-equilibrium limit in the scenario $\J = \Xi$ and $\mathcal{L}$ independent from $\Xi$ can be found in~\cite{krotov2021large, tang2021remark, fanaskov2024associativememorydeadneurons}. 

\subsection{Synaptic dynamics and Hebbian theory}\label{sec:Synaptic_discussion}
In the present work, the network components consist of neurons and memories.  
As highlighted previously in Sec.~\ref{sec:memory_generating_example}, in Krotov's previous work, memories act as synapses within this framework, which differs from the traditional interpretation of synapses as the product of memories, as Hebb's rule would prescribe.  
We have shown that by allowing the generating function to depend on memories, it is possible to recover the conventional interpretation (see Eq.~\ref{eq:almost_Hebb}). Here, we focus on the connections of our framework with Hebbian learning that prescribes recipes for the synaptic matrix that are used in Hopfield-like models.

\subsubsection{Connection to Hopfield model theory}\label{sec:Hop}
The retrieval process in Hopfield-like models works as follows~\cite{AGS_prl, hopfield1982}: one starts from an initial configuration and then updates it with a stochastic criterion that minimizes the energy and satisfies the detailed balance condition. If the load of the network is not too high the configuration state converges to one of the stored memories, and these memories are binary vectors approximately orthogonal to each other by hypothesis.
It is required that the initial configuration is a noisy version of one such memory, i.e. the distance with one of the patterns should be substantially greater than those with the other. The whole dynamics consists in descending the energy landscape downhill the attraction basin of this pattern. \\
Moving to our case, we observe that the requirement~\eqref{eq:condition} is analogous to those needed in the retrieval process of Hopfield models.\\
Another interesting remark concerns the relation between the coupling matrix typically used in Hopfield model $\bb J$ and our memory matrix $\Xi$. The former is obtained from the memories through the Hebb rule, its element are related to the memories as
\begin{equation}
  J_{ij} = \frac{1}{K}\sum _\mu \xi _i ^\mu \xi _j ^\mu \ .
\end{equation}
In our case $K= N_h$, and we have already observed that $\bb J = \Xi ^ \top \Xi / N_h = \Xi ^2 / N_h$.
Furthermore, for every orthogonal matrix $U$, we still have $\bb J = (U \Xi)^2 / N_h$.\footnote{
   This also means that we can choose $U$ so that $U \Xi$ is a triangular matrix and write $\bb J$ as a product of triangular matrices. This means the representation $\bb J$ is invariant upon orthogonal transformation, then, we can rotate the neural variables to diagonalize it~\cite{Barra2018}. Finally, we mention that Cholensky decomposing $\bb J$ can help in computational studies.
   }\\
We highlight that, while the Hebbian rule builds a $N_v \times N_v$  matrix from which memory vectors cannot be unambiguously recovered, our method, based on the  $N_h \times N_v$ memory matrix explicitly stores memories.
Furthermore, when $N_h < N_v$ holds, the memory matrix requires less storage than the Hebbian matrix. 
A model in which $N_h < N_v$ is the traditional Hopfield network \cite{hopfield1982, AGS_prl}. Here, $N_h$ corresponds to the number of memory patterns $K$ and the network works properly when the load $\alpha = \lim _{N_v \rightarrow \infty} \frac{N_h}{N_v}$ is below the critical threshold  $\alpha_\text{critical} \simeq 0.138$.

\subsubsection{Recovering Hebbian learning with synaptic evolution}\label{sec:hebbian}
At this point, it also becomes important to verify that the proposed memory evolution is consistent with Hebbian learning~\cite{Attneave1950b}.
In section~\ref{sec:memory_generating_example}, we have recovered the neural learning rules discussed in~\cite{lotito2024learningassociativenetworkspavlovian, Agliari2023}, that have been used to model pavlovian classical conditioning.\\
The same papers propose a synaptic dynamics that converges to the Hebbian prescription for the coupling matrix. This dynamics can be rewritten in our notation as
   \begin{equation}\label{eq:mem4pavlov} 
     \frac{\d}{\d t} ({\Xi ^\top \Xi }) = -    \Xi ^{\top} \Xi  + N_h \Xi ^{\prime \top} \text{diag} (\bb p)  \Xi ^\prime . 
   \end{equation} 
This equation depicts the convergence to the generalized Hebbian coupling matrix~\cite{lotito2024learningassociativenetworkspavlovian}, $\bb J =  \Xi^ {\prime \top  }\text{diag}(\bb p) \Xi ^ \prime$, or $ J_{ij} = \sum _\mu p _\mu \xi _i ^{\prime \mu} \xi _j ^{\prime \mu}$.
Our dynamics does not lead to the generalized Hebbian coupling matrix, but rather to the traditional Hebbian prescription $\bb J =  \Xi ^ {\prime \top  } \Xi ^{\prime} / N$. This is because in our case each pattern consolidates the corresponding memory matrix row, while $\bb p$ helps to quantify how fast each row converges (Sec.~\ref{sec:hierarchical}).\\
The learning rule \eqref{eq:mem4pavlov} can thus be obtained \textit{only} when $ p_\mu = 1/N_h $ for each $\mu=1,\dots, N_h$: in this case $  N_h \text{diag} (\bb p) = \Id$.
The conditions for convergence~\eqref{eq:conv_mem} are the same as discussed in Sec.~\ref{sec:hierarchical}. For instance, we need $\beta \gg 1 $, Eq.~\ref{eq:condition},  $\bb \xi ^{(0)} _\mu\approx \bb \xi ^\prime_\mu $ for each $\mu$, we also need the convergence time scales  $\{ \frac{p_\mu w_\mu}{\tau _\Xi} \doteq \tau_\mu \}_{\mu = 1, \dots, N_h}$ to be much bigger than the frequency of clamping events, such as in \cite{lotito2024learningassociativenetworkspavlovian}).
Under these condition we have $ \Xi _\mu   (t) = e^{-t/\tau_\mu} \Xi _\mu ^{(0)}   + (1 - e^{-t/\tau_\mu})  \Xi ^\prime _\mu $. We can now look at the evolution
\begin{align}\label{eq:coupling_expansion}
(\Xi^\top \Xi)_{ij}(t) =\sum_{\mu=1}^{N_h} \left\{  \left[  e^{-t/\tau_\mu} \xi_i^{(0)\mu} + (1-e^{-t/\tau_\mu})\xi_i^{\prime\mu}\right] \right.  \notag \\
\times \left. \left[e^{-t/\tau_\mu} \xi_j^{(0)\mu} + (1-e^{-t/\tau_\mu})\xi_j^{\prime\mu}\right] \right\}
\end{align}
Taking the limit as $t \to \infty$, we obtain the traditional Hebbian prescription:
\begin{align}\label{eq:coupling_limit}
\lim_{t \to \infty} (\Xi^\top \Xi)_{ij}(t) = \sum_{\mu=1}^{N_h} \xi_i^{\prime\mu} \xi_j^{\prime\mu} = (\Xi^{\prime\top} \Xi^{\prime})_{ij} \, .
\end{align}


\subsection{Eigenvalue problems in multi scale dynamics}\label{sec:connection_dyn}
In this section, we draw some connections with both dynamical system theory and random matrix theory. \\
We consider the simple case where $\L_h = \frac{1}{2}\sum_\mu h_\mu ^2$ and $\L_v = \frac{1}{2}\sum_i v_i ^2$ (quadratic terms in the energy are ubiquitous in science), both $\bb f$ and $\bb g$ are the identity function.\\
In the synaptic time scale, the algebraic equation for the visible neurons is $\bb v = \Xi ^\top \Xi \bb  v$, thus $\bb  v$  solves an eigenvalue problem. We can solve it and rotate the neural variables to diagonalize the system, $\tilde{\bb v} = \text{diag}(\bb\lambda) \tilde{\bb v}$.\\ %
We now move to the visible neuron time scale, the evolution of the visible neurons is
\begin{equation}  \label{eq:v_rm}
\tau _v \dot{ \bb v } = \left( \Id - \Xi ^\top \Xi \right) \bb v \, .  
\end{equation}
The system is linear and can be approached by standard methods.\\
However, the situation changes when considering the widely used hyperbolic tangent as activation functions,
\begin{align*}
\bb g(\bb v) &= \tanh(\bb v) \quad \text{(applied elementwise)} \\
\bb f( \bb h) &= \tanh(\bb h) \quad \text{(applied elementwise)}
\end{align*}
This choice corresponds to the generating functions $\L_h = \sum_\mu \ln (\cosh h_\mu )$, $\L_v =\sum_i \ln (\cosh v_i )$, and the visible neurons evolution reads as
\begin{align}
\tau_v \dot{\bb v} = \bb \psi_v(\bb v) =& -\bb v + \bb I + \Xi^\top \bb f(\Xi \bb g^\top + \bb b)^\top \notag  \\
=& -\bb v + \bb I + \Xi^\top \tanh(\Xi \tanh(\bb v)^\top + \bb b)^\top \, , \label{eq:vdyn_tanh} 
\end{align}
A starting point to study such non-linear evolution is to verify the conditions for the
Hartman-Grobman theorem to hold~\cite{arrowsmith1992dynamical}. This theorem connects the behavior of a non-linear system around its fixed points to its linear approximation. It states that for hyperbolic fixed points small higher-order perturbations to the system do not change the local structure near these points. For a fixed point $\dot{\bb v} =0$ to be hyperbolic, we require that the Jacobian $ \left( \frac{\d \bb \psi_v}{\d \bb v} \right) $ has all eigenvalues with real part different from zero.
Thus, the eigenvalues of the Jacobian play an important role in our problem.
To compute the Jacobian, we
first observe that $\frac{\partial}{\partial v_j}(-v_i + I_i) = -\delta_{ij}$.
We move to the term $\Xi^\top \bb f(\Xi \bb g^\top + \bb b)^\top$ and use the chain rule:
\begin{align*}
\frac{\partial}{\partial v_j} \left[\Xi^\top \bb f(\Xi \bb g^\top + \bb b)^\top\right]_i &= \sum_\mu \Xi^\top_{i\mu} \frac{\partial}{\partial v_j} f_\mu(\sum_k \Xi_{\mu k}g_k(v_k) + b_\mu) \\
&= \sum_\mu \Xi^\top _{i\mu} f'_\mu(\sum_k \Xi_{\mu k}g_k(v_k) + b_\mu) \sum_k \Xi_{\mu k}\frac{\partial g_k(v_k)}{\partial v_j} \\
&= \sum_\mu \Xi^\top_{i\mu} [1-\tanh^2(\sum_k \Xi_{\mu k}\tanh(v_k) + b_\mu)] \Xi_{\mu j}[1-\tanh^2(v_j)] \, .
\end{align*}
Here we used $\frac{\partial}{\partial v_j} g_i (v_i)= \frac{\partial}{\partial v_j}\tanh(v_i) = \delta_{ij} [ 1 - \tanh ^2(v_i)]$.
Therefore, the Jacobian can be written in matrix form as:
\begin{equation}
\frac{\partial \bb \psi_v}{\partial \bb v} = -\Id + \Xi^\top \left[\Id- \text{diag} \left(\tanh^2(\Xi \tanh(\bb v)^\top + \bb b)\right)\right] \Xi \left[\Id-\text{diag}\left(\tanh^2(\bb v)\right)\right]
\end{equation}
If the biases are set to zero $\bb b = \bb I = \bb 0$ , a fixed point is $\bb v^* = \bb 0$, we have:
\begin{align*}
\tanh(\bb v^*) &= \bb 0 \\
\Id- \text{diag}\left[ \tanh^2(\bb v^*)\right] &= \Id\\
\Id- \text{diag}\left[\tanh^2(\Xi \tanh(\bb v^*)^\top ) \right]&=  \Id
\end{align*}
Thus, at this fixed point, the Jacobian reduces to:
\begin{equation}
\left.\frac{\partial \bb \psi_v}{\partial \bb v}\right|_{\bb v^*} = -\Id + \Xi^\top  \Xi \, .
\end{equation}
Thus, in the zero bias case, linearizing Eq.~\ref{eq:vdyn_tanh} around the zero fixed point recovers Eq.~\ref{eq:v_rm}. The condition under which the fixed points of the system are hyperbolic is that 
\textit{(i) all eigenvalues of the matrix $\Xi ^\top \Xi$ have a real part different from $1$}.\\
We continue this analysis by considering the random setting in which the memories are drawn from a normalized Gaussian distribution $\xi ^\mu _i \sim \mathcal{N}(0, N_h^{-1}), \ \forall i,\mu$. Analogous settings are frequent in the study of spin glasses \cite{mezard_parisi_virasoro_1987}.
The matrix $\Xi ^\top \Xi$ is a Wishart matrix whose eigenvalues spectrum follows the Marchenko-Pastur distribution for large $N_v$ and fixed $\alpha = \lim _{N_v \rightarrow \infty} \frac{N_h}{N_v}$~\cite{livan2018introduction}.\footnote{In the typical statistical mechanics setting $\alpha$ corresponds to the load of the network, and the limit we are taking is the thermodynamic limit~\cite{barra2018phase}.}\\
Condition \textit{(i)} is automatically satisfied if the smallest eigenvalue $\lambda_{\min}$ is such that $\lambda_{\min} > 1$.
In this random setting, in the thermodynamic limit, $\lambda _{\min} = (1 - \sqrt{\alpha ^{-1}})^2$. 
Thus, in this limit, provided that $\alpha < 1/4$, the equilibria of Eq.~\ref{eq:vdyn_tanh} system are hyperbolic fixed points. A slightly more general discussion can be found in Appendix~\ref{app:HG}. \\

\section{Final remarks and conclusion}\label{sec:conclusions}
In this work, we have considered an artificial neural network under the assumption of well-separated timescales. In this network, the neural activation functions are linked to the energy of the whole system~\cite{krotov2021large, krotov2021hierarchical}. \\
In this dynamical picture of associative networks, the energy and the activations are obtained from a generating function through Legendre transformation and derivation w.r.t. neural variables, respectively.

In this paper, we have established the theoretical foundations of Krotov's picture of the dynamics of associative networks~\cite{krotov2021large}.
We implemented learning in this framework and used results from various branches of mathematics to connect it with Hebbian theory. \\
First, we allowed the generating function to depend on the memory matrix in order to recover the traditional interpretation of synaptic strength and to reproduce the Hebbian prescription for couplings within our framework. 
Moreover, we showed that this ``Hebbian rule'' can be recovered as a stable point of our synaptic dynamics. 

In conclusion, we believe that our work lays a solid mathematical foundation and offers significant theoretical advancements in associative network theory,  broadening its relevance to various disciplines.

\section*{Acknowledgments}
The author acknowledges Adriano Barra, Gaia Marangon, Michael Johnston and Dmitry Krotov for valuable discussions.

\printbibliography

\newpage
\appendix
\section{Further connections to dynamical systems theory} \label{app:HG}
We consider the case $\J = \Xi$ and $\mathcal{L}$ independent from $\Xi$, the neural dynamics~\eqref{eq:vxidyn}  corresponds to the dynamical system $\tau _v \dot{ \bb v } = \bb \psi_v(\bb v)$ with $\bb \psi_v(\bb v) = - \bb v +\bb I + \Xi^\top \bb f( \Xi\bb g ^{\top}  + \bb b ) ^ \top$.
To study this system, we compute the Jacobian of $\bb \psi _v$,\\
\begin{equation}\label{eq:jacphi}
  \left( \frac{\d \bb \psi _v}{\d \bb v} \right) = - \Id + \Xi ^\top  \mathbf {H}_v \mathbf {H}_h \Xi \, ,
\end{equation}
where we recognize the Hessians $\mathbf {H}_v, \ \mathbf {H}_h $ of $\L _v, \ \L _h$ respectively; they are both positive definite since the generating functions are convex.\footnote{Moreover, the natural assumption of linearly independent memories $\{ \xi \}$ leads to a full rank matrix $\Xi$.}
This allows us to compute the eigenvalues of the Jacobian~\eqref{eq:jacphi} and determine whether the Hartman-Grobman theorem applies to our system \cite{arrowsmith1992dynamical}. According to this theorem, small higher-order perturbations do not alter the local structure near hyperbolic fixed points. A fixed point \( \dot{\bb v} = 0 \) is hyperbolic if the Jacobian \( \frac{\d \bb \psi_v}{\d \bb v} \) has no eigenvalues with zero real part.
In our case, this corresponds to requiring that $\Xi ^T \mathbf {H}_v \mathbf {H}_h \Xi $ has no eigenvalue with real part equal to one, and the theorem would ensure that in the neighborhoods of these fixed points the non-linear system $\tau _v \dot{ \bb v } = \bb \psi_v(\bb v)$ behaves like the linearized system $\tau _v \dot{ \bb v } = \left( \frac{\d \bb \psi _v}{\d \bb v} \right) \bb v  $.\\ 
To conclude, we mention that the recent work by Fanaskov and Oseledets is entirely dedicated to exploring this connection 
\cite{fanaskov2024associativememorydeadneurons}.

\section{Synaptic dynamics analysis}\label{app:poc}
As a proof of concept we assume zero bias \(\Lambda = 0\), and numerically simulate the learning scenario depicted in the main text:
we start from a set of old memories \(\{ \bb \xi^{(0)} \}\) that form the memory matrix \(\Xi^{(0)}\). We update these memories based on observed patterns \(\{ \bb \xi^\prime \}\), which correspond to the possible values visible neurons can assume.  These are patterns that be grouped into a new memory matrix \(\Xi^\prime\).\\
We consider typical assumptions of statistical mechanics of associative networks~\cite{mezard2009information, Loureiro2021, barramnist2021}. 
In particular, we assume that each observable pattern $\bb \xi _\mu ^\prime$ is a perturbed version of the corresponding memory $\bb \xi ^{(0)}_\mu$. A valid choice, that we adopt in our proof of concept (Sec.~\ref{app:poc_num}), could be
   \begin{equation}
   \xi^\prime_{\mu i} = \xi^{(0)}_{\mu i} + \eta_{\mu i}, \quad \eta_{\mu i} \sim \mathbb{P}_\eta \, ,
   \end{equation}
where the noise terms are independent across both indices and are characterized by the zero-mean distribution $\mathbb{P}_\eta$. While, for the initial patterns, Eq.~\ref{eq:condition} holds.
\subsection{Theoretical Analysis of Memory Evolution}\label{app:poc_th}
Starting from the evolution equation for each memory row $\mu$,
\begin{equation}\label{eq:evolution}
    \tau_\Xi \frac{\d}{\d t} \Xi_\mu = w_\mu (-\Xi_\mu + \bb v),
\end{equation}
where $\bb v$ is clamped to pattern $\bb \xi^\prime_\nu$ with probability $p_\nu$. For large $\beta$, when pattern $\nu = \mu$ is shown, $w_\mu \approx 1$, while for $\nu \neq \mu$, $w_\mu \approx 0$. This leads to the effective evolution equation,
\begin{equation}\label{eq:effective}
    \tau_\Xi \frac{\d}{\d t} \Xi_\mu = p_\mu (-\Xi_\mu + \bb \xi^\prime_\mu),
\end{equation}
where we account for the probability $p_\mu$ of showing pattern $\mu$. The solution to this differential equation reads as
\begin{equation}\label{eq:solution}
    \Xi_\mu(t) = e^{-t/\tau_\mu} \Xi_\mu^{(0)} + (1 - e^{-t/\tau_\mu}) \bb \xi^\prime_\mu,
\end{equation}
with characteristic time scale
\begin{equation}\label{eq:tau}
    \tau_\mu = \frac{\tau_\Xi}{p_\mu}.
\end{equation}
We study the convergence of each memory row toward its target pattern by tracking their similarity over time. A natural measure for this is the cosine similarity, which for two vectors is defined as the ratio between their dot product and the product of their norms. The cosine similarity between the evolving memory $\Xi_\mu(t)$ and its target $\bb \xi^\prime_\mu$ reads as
\begin{equation}\label{eq:cosine}
    \text{sim}_\mu(t) = \frac{\Xi_\mu(t) \cdot \bb \xi^\prime_\mu}{\|\Xi_\mu(t)\| \|\bb \xi^\prime_\mu\|}.
\end{equation}
We denote the initial cosine similarity between memory and target pattern as $\alpha_\mu$,
\begin{equation}\label{eq:alpha_def}
    \alpha_\mu = \frac{\Xi_\mu^{(0)} \cdot \bb \xi^\prime_\mu}{\|\Xi_\mu^{(0)}\| \|\bb \xi^\prime_\mu\|}.
\end{equation}
Given the normalization condition $\|\Xi_\mu^{(0)}\|^2 = N_v \simeq \|\bb \xi^\prime_\mu\|^2 $,  we express $\Xi_\mu^{(0)} \cdot \bb \xi^\prime_\mu \simeq \alpha_\mu N_v$. 
Using this approximation, we compute the numerator of the cosine similarity,
\begin{equation}\label{eq:numerator}
    \Xi_\mu(t) \cdot \bb \xi^\prime_\mu = N_v\left[e^{-t/\tau_\mu}\alpha_\mu + (1 - e^{-t/\tau_\mu})\right],
\end{equation}
and the norm of the evolving memory,
\begin{equation}\label{eq:norm_evolving}
    \|\Xi_\mu(t)\|^2 = N_v\left[e^{-2t/\tau_\mu} + (1 - e^{-t/\tau_\mu})^2 + 2e^{-t/\tau_\mu}(1-e^{-t/\tau_\mu})\alpha_\mu\right].
\end{equation}
Combining these results, the cosine similarity evolves as
\begin{equation}\label{eq:sim_final}
    \text{sim}_\mu(t) = \frac{e^{-t/\tau_\mu}\alpha_\mu + (1 - e^{-t/\tau_\mu})}{\sqrt{e^{-2t/\tau_\mu} + (1 - e^{-t/\tau_\mu})^2 + 2e^{-t/\tau_\mu}(1-e^{-t/\tau_\mu})\alpha_\mu}}.
\end{equation}
We appreciate that in the limit $t \to \infty$, the similarity approaches unity. Moreover, since $(\Xi_\mu^{(0)} \cdot \bb \xi^\prime_\mu)\propto N_v$, the similarity evolution does not depend on $N_v$.

\subsection{Numerical Analysis of Memory Evolution}\label{app:poc_num}
\begin{figure}[t]
\centering
\includegraphics[width=0.7\textwidth]{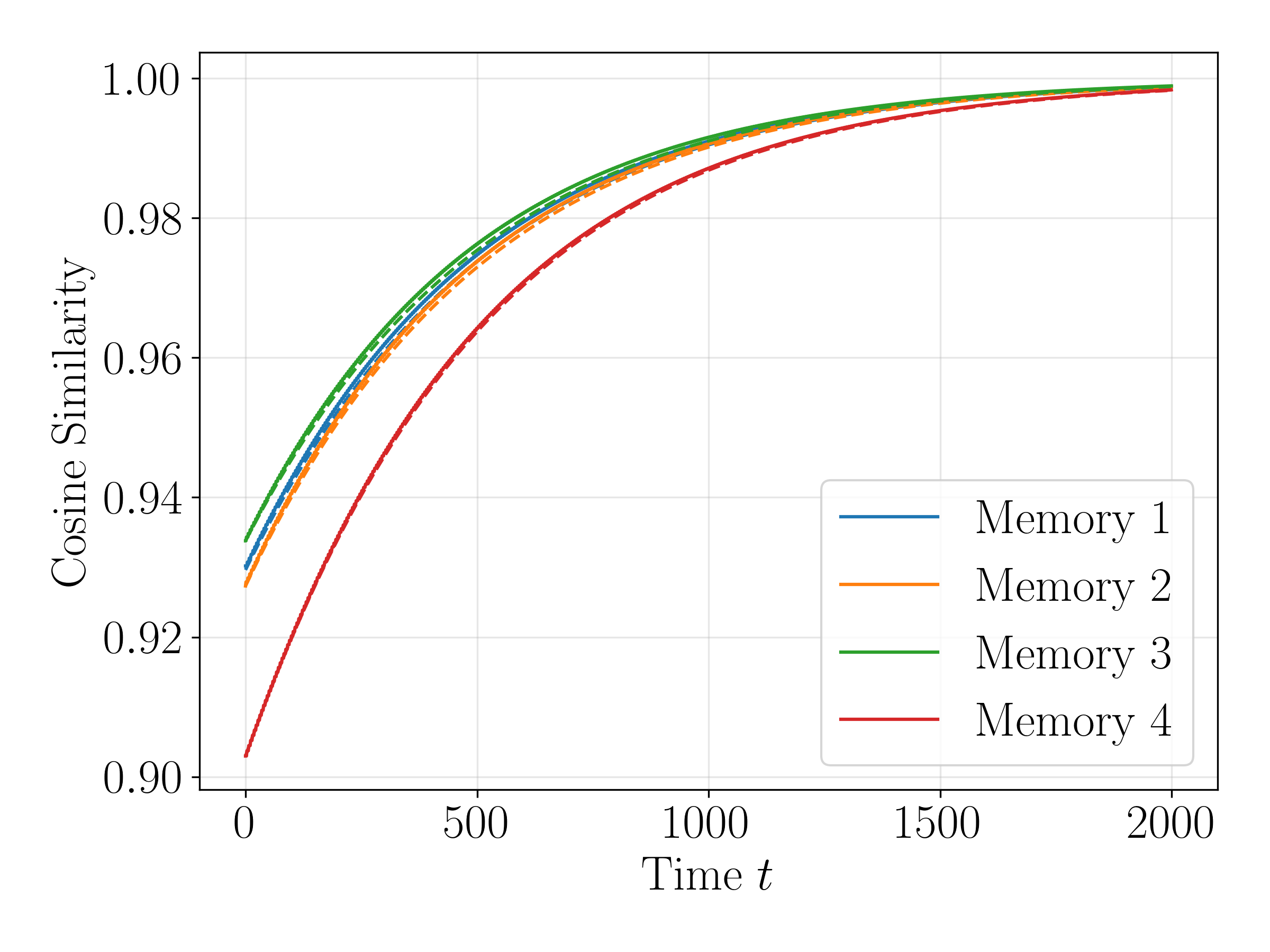}
\caption[Memory dynamics theoretical comparison]{Evolution of cosine similarity between each memory row $\Xi_\mu(t)$ and its target pattern $\bb \xi^\prime_\mu$ as described by Eq.\ref{eq:cosine}. Solid lines represent numerical simulations obtained by simulating the synaptic dynamics (see Eq.\ref{eq:memdyn_el}) with parameters $N_v=50$, $N_h=4$, $\tau_\Xi=250$, $\beta=2$, and noise level $\sigma=0.4$. Initial memories are constructed to be perfectly orthogonal through a Gram-Schmidt process and scaled to norm $\sqrt{N_v}$. Dashed lines show the theoretical prediction from Eq.\ref{eq:sim_final}, where the characteristic time scales $\tau_\mu$ follow from Eq.\ref{eq:tau} with uniform probabilities $p_\mu=1/N_h$. The perfect agreement confirms our analytical understanding of the memory evolution process for a wide range of parameter values.}
\label{fig:app_poc}
\end{figure}

For our proof of concept, each entry of the observed patterns includes an additive independent Gaussian noise:
   \begin{equation}
   \xi^\prime_{\mu i} = \xi^{(0)}_{\mu i} + \eta_{\mu i}, \quad \eta_{\mu i} \sim \mathcal{N}(0,\sigma^2)
   \end{equation}
where the noise terms are independent across both indices. While, for the initial patterns, it holds the perfect orthogonality condition
\begin{equation}\label{eq:condition_perfect}
  \frac{\bb \xi^{(0)}_\mu \cdot \bb \xi^{(0)}_\nu}{\| \bb \xi^{(0)}_\mu \| \| \bb \xi^{(0)}_\nu \|} = \delta_{\mu \nu} 
\end{equation}
for each \(\mu, \nu \in \{1, \dots, N_h\}\).
Numerically, we follow a Gram-Schmidt process to ensure orthogonality, and then we scale the vectors so that their norms are \( \| \bb \xi^{(0)}_\mu \|^2 = N_v\). Consequently, since visible neurons are always clamped to one of the $\bb \xi ^ \prime$ vectors, it holds that \(\| \bb v \| ^2 \simeq N_v\).
We illustrate that we obtain the desired outcome $\Xi^{(0)} \longrightarrow \Xi^{(\infty)} = \Xi^\prime.$ in Fig.\ref{fig:app_poc}.
Strikingly, the result is in perfect agreement with the theoretical prediction given by Eq.~\ref{eq:sim_final}.
The numerical simulations have been performed with a moderate system size $N_v=50$ and $N_h=4$ memories. The time evolution spans $T=2000$ time steps, with a memory time scale $\tau_\Xi=250$. We set a noise level $\sigma=0.4$ and an inverse temperature $\beta=2$ to match the theoretical predictions for large $\beta$.\footnote{Setting $\beta > 3$ results in the desired behavior in all tested settings.} While we present results for this specific set of parameters, we have verified the robustness of our findings across different parameter ranges. The complete Python implementation is available at~\cite{memoryGithub}~\footnote{\url{https://github.com/danielelotito/softmax-synapses}}.

\end{document}